\documentclass[aps,pre,showpacs,amsmath,amssymb,amsfonts,lengthcheck,twocolumn,longbibliography,superscriptaddress]{revtex4-2}

\usepackage{changes} 

\usepackage{graphicx}
\usepackage{subfigure}
\usepackage{amsthm}
\usepackage{amsmath}
\usepackage{verbatim}
\usepackage{dcolumn}
\usepackage{bm}
\usepackage{color}
\usepackage[colorlinks=true,citecolor=blue,linkcolor=blue,urlcolor=blue]{hyperref}%
\usepackage{xcolor}
\usepackage{dsfont}
\usepackage{longtable}
\usepackage{hyperref}
\allowdisplaybreaks

\newcommand{\bla}{bla\\bla\\bla\\bla\\bla}

\begin{document}

\title{Magnetocaloric effect for a $Q$-clock type system}

\author{Michel Aguilera}
\email{michel.aguilera@usm.cl}
\affiliation{Instituto de Física, Pontificia Universidad Católica de Valparaíso, Casilla 4950, Valparaíso 2373223,
Chile}
\affiliation{Departamento de F\'isica, Universidad T\'ecnica Federico Santa Mar\'ia, Av. Espa\~na 1680, Valpara\'iso 2390123, Chile}
\author{Sergio Pino-Alarc\'{o}n}
\affiliation{Departamento de F\'isica, Universidad T\'ecnica Federico Santa Mar\'ia, Av. Espa\~na 1680, Valpara\'iso 2390123, Chile}
\author{Francisco J. Peña}
\email{francisco.penar@usm.cl}
\affiliation{Departamento de F\'isica, Universidad T\'ecnica Federico Santa Mar\'ia, Av. Espa\~na 1680, Valpara\'iso 2390123, Chile}
\affiliation{ Millennium Nucleus in NanoBioPhysics (NNBP), Av. Espa\~na 1680, Valpara\'iso 2390123, Chile}
\author{Eugenio E. Vogel}
\affiliation{Departamento de Ciencias F\'isicas, Universidad de La Frontera, Casilla 54-D, Temuco, Chile}
\affiliation{Facultad de Ingenier\'{\i}a, Universidad Central de Chile, Santiago 8330601, Chile }
\author{Natalia Cortés}
\affiliation{Instituto de Alta Investigación, Universidad de Tarapacá, Casilla 7D, Arica, Chile}
\author{Patricio Vargas}
\affiliation{Departamento de F\'isica, Universidad T\'ecnica Federico Santa Mar\'ia, Av. Espa\~na 1680, Valpara\'iso 2390123, Chile}

\date{\today}

\begin{abstract}
In this work, we study the magnetocaloric effect (MCE) in a working substance corresponding to a square lattice of spins with $Q$ possible orientations, known as the ``$Q$-state clock model". When the $Q$-state clock model has $Q\geq 5$ possible configurations, it presents the famous Berezinskii–Kosterlitz–Thouless (BKT) phase associated with vortices states. We calculate thermodynamic quantities using Monte Carlo simulations for even $Q$ numbers, ranging from $Q=2$ to $Q=8$ spin orientations per site in a lattice. We use lattices of different sizes with $L\times L = 8^{2}, 16^{2},  32^{2}, 64^{2}, \text{and}\ 128^{2}$ sites, considering free boundary conditions and an external magnetic field varying between $B = 0$ and $B=1$ in natural units of the system. By obtaining the entropy, it is possible to quantify the MCE through an isothermal process in which the external magnetic field on the spin system is varied. In particular, we find the values of $Q$ that maximize the MCE depending on the lattice size and the magnetic phase transitions linked with the process. Given the broader relevance of the $Q$-state clock model in areas such as percolation theory, neural networks, and biological systems, where multi-state interactions are essential, our study provides a robust framework in applied quantum mechanics, statistical mechanics and related fields. 

\end{abstract}

\maketitle

\section{Introduction}
\label{intro}

Caloric phenomena form a foundational aspect of material physics, crucial in identifying viable substitutes for the toxic gases currently employed in conventional refrigeration systems \cite{reis2020caloric,ram2018review}. Significant temperature variations driven by caloric processes have been observed in different materials following adiabatic changes in applied hydrostatic pressure (barocaloric effect) \cite{cirillo2022cooling,li2019colossal,matsunami2015giant,ma2017barocaloric}, mechanical stress (elastocaloric effect) \cite{xiao2015elastocaloric,chen2021elastocaloric,cong2019colossal,chauhan2015elastocaloric,chauhan2015review}, electric field (electrocaloric effect) \cite{mischenko2006giant,liu2016direct,correia2013electrocaloric,jia2012solid}, and external magnetic field (magnetocaloric effect, MCE) \cite{tishin2016magnetocaloric,de2010theoretical,pecharsky1999magnetocaloric,franco2018magnetocaloric}. Although the underlying nature and physical interpretation of each phenomenon differ, they share a common objective: maximizing entropy changes in response to variations of the control parameter governing the thermodynamic process \cite{reis2020magnetocaloric,reis2012oscillating,reis2011oscillating,reis2014diamagnetic,martinez2023caloric,negrete2018magnetocaloric,negrete2018magnetocaloric2}. 

Systems exhibiting phase transitions, in particular, achieve maximum entropy changes near where these transitions are occurring \cite{tegus2002magnetic,tishin1999magnetocaloric,de2008magnetocaloric,franco2009magnetocaloric}. Among these, magnetic materials stand out due to their distinct magnetic order or ground state, which is heavily influenced by the energy contributions within the material \cite{halder2010magnetocaloric,shamberger2009hysteresis,zhang2008magnetostructural}. Here we concentrate on magnetic systems that undergo phase changes with increasing temperature, specifically those that alter their magnetic phase. A notable model for exploration is the $Q$-state clock model, a discrete variant of the renowned 2D XY model \cite{tobochnik1982properties,li2020critical,miyajima2021machine,lupo2017approximating}, widely studied for illustrating the Berezinskii–Kosterlitz–Thouless (BKT) transition in frustrated quenched disordered phases \cite{ilker2014odd,chatterjee2020ordering,ali2024quantum}. The $Q$-state clock model serves as a classical Heisenberg spin model with pronounced planar anisotropy, useful in replicating material thermodynamics \cite{negrete2021short, negrete2018entropy,aguilera2022otto}.

Phase transitions in the $Q$-clock state model can be identified via the maxima observed in the specific heat as a function of temperature. Each maximum signifies there is a "critical temperature." In scenarios devoid of an external magnetic field, studies have demonstrated \cite{elitzur1979phase,cardy1980general,frohlich1981kosterlitz,ortiz2012dualities,kumano2013response,kim2017partition} that for $Q \geq 5$, the specific heat exhibits two maxima. The initial maximum pertains to the transition from a ferromagnetic phase (FP) to a BKT phase, followed by a transition from BKT to a paramagnetic disordered phase (PP) \cite{miyajima2021machine}.

Due to these dual phase transitions, the $Q$-clock state model is particularly intriguing for caloric phenomenon studies. This model prompts several critical questions: What is the optimal value of $Q$ that maximizes entropy variation and, consequently, the MCE? Which magnetic phases enhance the effect most effectively? Are the results consistent as the lattice size increases?

A major technical challenge with this model revolves around scaling accessible states, which scales approximately as $\sim Q^{L^{2}}$, where $N=L \times L$ defines the lattice size with $N$ number of sites. Consequently, exact computations using the canonical partition function entail substantial expenses, thus restricting exact computations to smaller lattices. Mean Field approximation and Monte Carlo simulations allow us to address large lattice size challenges. Monte Carlo simulations are often preferred for their accuracy in representing systems with intricate interactions and fluctuations, which the mean-field approach may overlook due to its averaging assumptions. Consequently, mean-field theory may not aptly describe short-range interactions like those in the systems examined herein.

Experimental validation of the $Q$-state clock model is complex due to its theoretical nature. Nonetheless, experimental scenarios reflecting behaviors predicted by this model, especially considering phase transitions, exist, notably in magnetic thin films and two-dimensional magnets \cite{li2020critical}, liquid crystals \cite{tuan2022binder}, and cold atom systems \cite{huang2018clean}.

This work investigates the MCE in the $Q$-clock model, utilizing Monte Carlo simulations to derive the model's thermodynamic properties for large lattice sizes. The analyzed $Q$ values are even numbers ($Q = 2, 4, 6, 8$), facilitating exploration into how the spin orientation given by $Q$ numbers influences the magnetocaloric effect across variations of thermal and magnetic conditions. We examine the magnetocaloric effect's reliance on these $Q$ values, focusing on entropy, specific heat, internal energy, and magnetization responses near phase transitions. This study elucidates optimal spin configurations to maximize the MCE in discrete multi-state systems.

Our work is structured as follows: Section \ref{Spin Model} introduces the $Q$-state clock model, detailing its Hamiltonian and the relevant thermodynamic quantities. Section \ref{Monte Carlo Simulations} describes the Monte Carlo simulation techniques used to derive thermodynamic properties, including detailed explanations of lattice sizes, boundary conditions, and sampling methodologies. Section \ref{Analysis of Thermodynamic Quantities} examines the calculated thermodynamic properties, such as specific heat, internal energy, and entropy, for various values of $Q$. Section \ref{Caloric Phenomena} explores the magnetocaloric effect and outlines the methodology for its calculation. Section \ref{Resultssec} presents our findings, highlighting the optimal conditions and $Q$ values that maximize the effect. Finally, Section \ref{Conclusec} summarizes the key findings of our research.

\section{Spin Model}\label{Spin Model}

\subsection{$Q$-state clock model}

\begin{figure}
	\subfigure[]{
		\includegraphics[width=.3\textwidth]{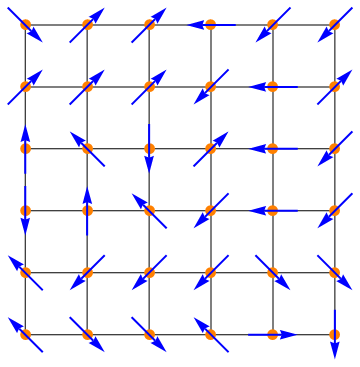}
	}
	\hspace{5mm}
	\subfigure[]{
		\includegraphics[width=.3\textwidth]{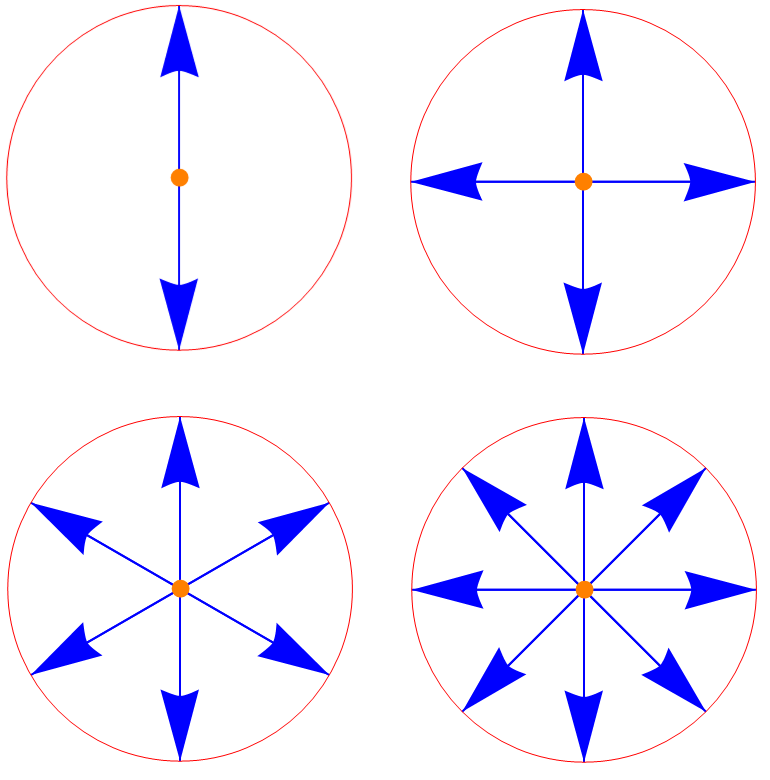}
	}
    \caption{(a) Schematic representation for a square lattice of size $6 \times 6$ with spin orientations corresponding to $Q=8$. (b) $Q$-clock model for $Q=2, 4, 6$ and $8$ states. Orange dots indicate sites, and blue arrows the possible spin orientation at each site with spin vector $\vec{S_{i}}$.}
    \label{modelfigure}
\end{figure}


The system under study corresponds to the $Q$-states clock model on a two-dimensional (2D) square lattice of dimensions $L \times L=N$ (with $N$ the total number of sites in the lattice). The local magnetic moment or ``spin'' $\vec{S}_i$ at site $i$ can point in any of $Q$ directions in a given plane, see Fig.\ \ref{modelfigure}. $\vec{S}_i$ is then a 2D vector, i.e. $\vec{S}_i=(\cos (\frac{2\pi}{Q} k), \sin (\frac{2 \pi}{Q} k))$, where $k=0,1,...Q-1$, with equal probability. The magnitude of $\vec{S}_i$ is chosen as the unity.

The isotropic Hamiltonian for such a system can be written as:
\begin{equation}
\mathcal{H} = - \sum_{\langle i,j \rangle} \mathcal{J} \vec{S}_i \cdot \vec{S}_j -  \sum_i \vec{B} \cdot \vec{S}_i   \, ,
\label{Clock}
\end{equation}
where $\mathcal{J}>0$ is the ferromagnetic exchange interaction to nearest neighbors; the sum runs over all pairs of nearest neighbors $\langle i,j \rangle$. $\vec{B}$ is an external magnetic field applied along one of the directions of the plane. We use $\mathcal{J} = 1$ in our calculations, thus all quantities are delivered in exchange units. 

To quantify the caloric effect, the entropy from the thermodynamics of the model must be obtained. To do that, we will analyze Monte Carlo simulations, which offer the possibility of obtaining thermal averages within a sample of the possible configurations of the system. The following section will detail the method employed, including the algorithm used in our computational calculations.

\section{Monte Carlo Simulations}\label{Monte Carlo Simulations}

The work presented herein will focus on numerical computations based on Monte Carlo (MC) simulations. A square lattice of size $L \times L$ is selected, with free boundary conditions imposed. A site is randomly chosen for visitation, and the energy cost, $\Delta$, associated with rotating the corresponding spin among $Q$ possible states is calculated. If the energy is diminished, the change in orientation is accepted; otherwise, the spin rotation is accepted only when $\exp(-\Delta/T) \le r$, where $r$ is a freshly generated random number uniformly distributed in the range [0,1]. This procedure conforms to the conventional Metropolis algorithm. A Monte Carlo step (MCS) is achieved after $N = L \times L$ spin-rotation attempts.


For each lattice size and $Q$ value, a sequence of temperatures has been established within the range [0.02, 4] in increments of 0.02 for each temperature. A total of $5\tau$ MCSs are performed: the initial $\tau$ MCSs are utilized to equilibrate the system at a fixed temperature $T$. In contrast, the subsequent $4\tau$ MCSs are employed to measure observables every 20 MCSs, achieving a cumulative total of $2 \times 10^5 = 200.000$ measurements. From now on, $\tau = 10^6$, unless otherwise specified for the remainder of this paper, this choice of $\tau$ yields stable results and corresponds with the analytical expressions obtained in previous studies for smaller lattices.

\subsection{Thermal averages}

The magnetization per site $\vec{M}$ is given by
\begin{equation}
	\vec{M} = \frac{1}{N} \sum _{i=1}^N \vec{S}_i,   \label{M}
\end{equation}
where $\vec{S}_i$ is the spin at site $i$ at a given time, $t$, and $N = L \times L $. In this case
$\vec{M}$ is a vector of two components $\vec{M}=(M_x,M_y)$. Normally, this vector's magnitude or absolute value is calculated as $|\vec{M}|=\sqrt{M_x^2+M_y^2}$. Then, the thermal average of $|\vec{M}|$ is $<|\vec{M}|>$ given by
\begin{equation}
	<|\vec{M}|> = \frac{1}{N_c} \sum _{i=1}^{N_c} \sqrt{M_x^2+M_y^2},  \label{abs(M)}
\end{equation}
with $N_c = 2 \times 10 ^5$ the number of configurations used to perform thermal averages.

Energy is the main quantity the Monte Carlo method uses to reach thermal equilibrium. Therefore, after  $\tau$ MCSs the internal energy $U$
can be obtained by averaging  the $N_c= 2 \times 10 ^5$ values for $E_k$, where $k$ runs over the accepted configurations after the Metropolis algorithm, namely
\begin{equation}
	U = <E> = \frac{1}{N_c} \sum _{k=1}^{N_c} E_k,  \label{ave(E)}
\end{equation}
where every spin configuration is separated from the next one by $20$ MCSs. The energy per site is then $U/N$.

The specific heat is then calculated as proportional to the fluctuations of the energy as follows (we use $k_{B}=1$ for simplicity):

\begin{equation}
	C  = \frac{\langle E^2 \rangle-\langle E \rangle ^2}{T^2},
\end{equation}
\begin{equation}
    C = \frac{1}{T^2}\left[\left( \frac{1}{N_c} \sum _{k=1}^{N_c} E _k^2 \right)- \left( \frac{1}{N_c} \sum _{k=1}^{N_c} E_k \right)^2\right]. \label{flucE}
\end{equation}
The entropy $S$ can be calculated by numerical integration of the specific heat over $T$ as 
\begin{equation}
	S(T,B) = S_{0} + \int \frac{C(T,B)}{T} dT.   \label{DeltaS}
\end{equation}


Our analysis determines the entropy at zero temperature under zero and non-zero magnetic field conditions by examining the energy degeneracy inherent at $(T=0)$. Without a magnetic field, $(Q)$ ferromagnetic spin configurations exist, each possessing identical energies, resulting in an entropy of $S_{0} = \ln Q$. Consequently, the entropy at any temperature $T$ with a zero magnetic field is given by $S(T,0) = \ln Q + \int \frac{C(T,0)}{T} dT$. Conversely, when the magnetic field $(B \neq 0)$, there is a unique ground state where all spins align uniformly with the field, leading to the constant $S_{0}$ going to zero.

\begin{figure}
	\subfigure[]{
		\includegraphics[width=.50\textwidth]{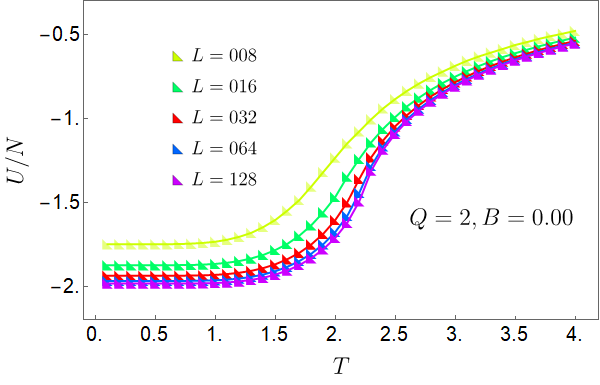}
	}
	\hspace{5mm}
	\subfigure[]{
		\includegraphics[width=.50\textwidth]{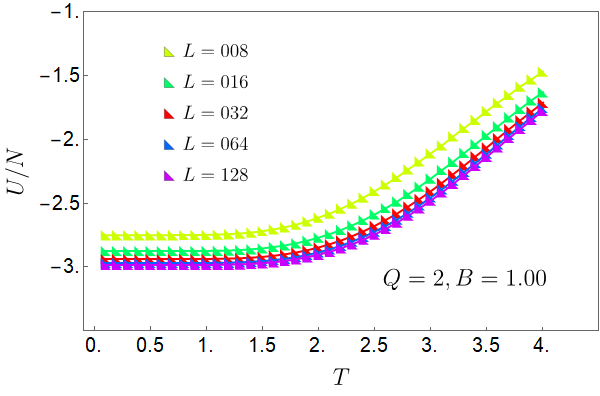}
	}
    \caption{(a) Normalized internal energy $U/N$ as a function of temperature for $Q=2$ and lattice sizes $8\times 8$, $16\times 16$, $32\times 32$, $64\times 64$ and $128\times128$ for an external magnetic field of (a) $B = 0$, (b) $B = 1$.}
    \label{sizelattice}
\end{figure}


\subsection{Selection of the lattice size}

Due to the scaling of the system's accessible states ($\sim Q^{L^{2}}$ ), it is necessary to set a criterion for a representative lattice size in the thermodynamic limit, where the results of the caloric studies are valid and closer to reality. For this purpose, we have analyzed the convergence in the internal energy $U$ for different values of the lattice size from our Monte Carlo simulations. For this purpose, we have decided to explore the internal energy of the system (normalized to the number of sites) as a function of $T$ for $Q=2$ and different lattice sizes, corresponding to $8\times 8$, $16\times 16$, $32\times 32$, $64\times 64$ and $128\times128$, and for magnetic field values of $B=0$ and the largest field that we will consider $B=1$. This is observed in Fig.~\ref{sizelattice}(a)-(b), where it can be seen that for a lattice size of $128\times 128$, the variation between their internal energy compared to the $64\times 64$ lattice is extremely small for both $B$ values. Therefore, a larger lattice, for example, $256\times 256$, would no longer show substantial differences in internal energy as in comparison to a $128 \times 128$ lattice, within the precision of Fig.~\ref{sizelattice}, which is enough for the scope of this paper. That is why our numerical thermodynamic limit will correspond in the following analysis to a lattice size of $128\times 128$.

\section{Analysis of Thermodynamic Quantities}\label{Analysis of Thermodynamic Quantities}

\begin{figure}
	\subfigure[]{
		\includegraphics[width=.50\textwidth]{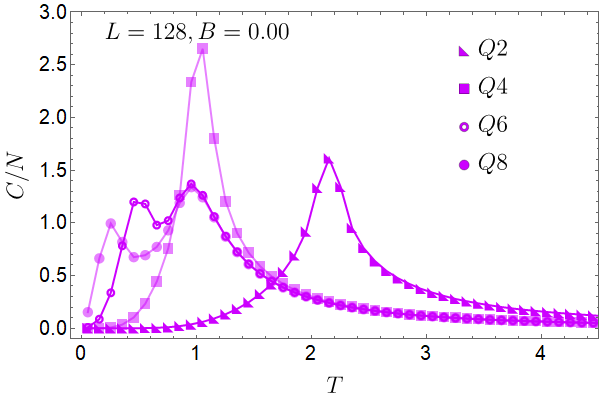}
	}
	\hspace{5mm}
	\subfigure[]{
		\includegraphics[width=.50\textwidth]{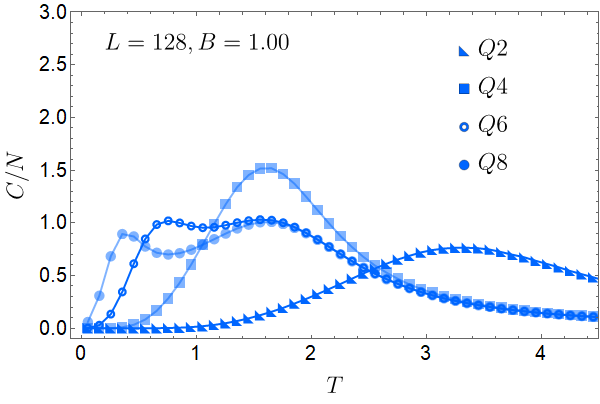}
	}
    \caption{Normalized specific heat $C/N$ as a function of temperature for even $Q$ values between $Q=2$ and $Q=8$ and a $128 \times 128$ lattice. (a) $B = 0$, (b) $B = 1$. The shift of the peaks to the right at $B=1$ is clearly seen for all $Q$.}
    \label{specificheatImage}
\end{figure}


We will begin by reviewing the thermodynamic results of the model, focusing first on the analysis of the specific heat for a  $128 \times 128$ lattice. This thermodynamic quantity for a field $B = 0$ and field $B = 1$ is shown in Fig.~\ref{specificheatImage}(a) and Fig.~\ref{specificheatImage}(b), respectively, for $Q=2,4,6,8$. Here, we observe two peaks for $Q \geq 6$, indicative of a double-phase transition. It is, therefore, the BKT phase that appears in these cases. It is also noticeable that the critical temperature of each transition increases as $B$ increases, and their associated peaks are less pronounced, as we can observe when comparing Fig.~\ref{specificheatImage} panel (a) with panel (b). 
This is because the external magnetic field favors ordered phases (FM and BKT) over disordered ones. 

\begin{figure}
	\subfigure[]{
		\includegraphics[width=.50\textwidth]{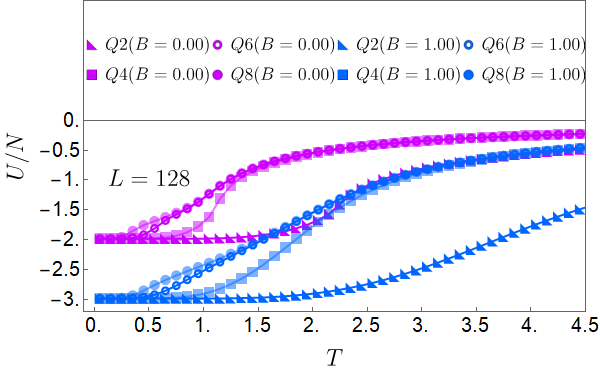}
	}
	\hspace{5mm}
	\subfigure[]{
		\includegraphics[width=.50\textwidth]{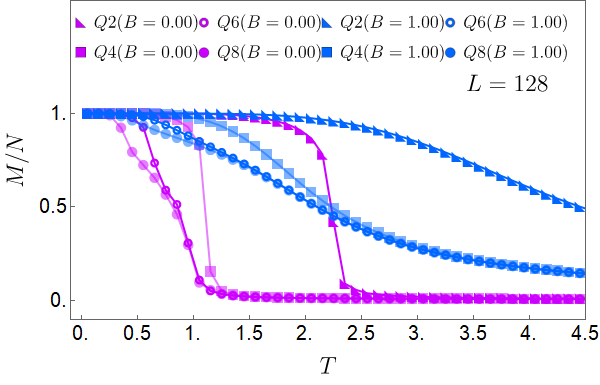}
	}
 \subfigure[]{
		\includegraphics[width=.50\textwidth]{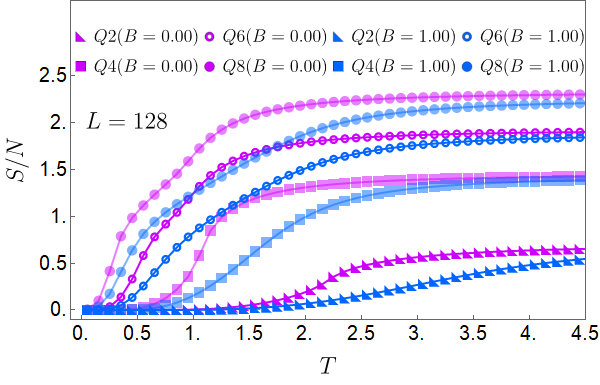}
	}
	\hspace{5mm}
    \caption{Normalized (a) internal energy, (b) magnetization, and (c) entropy as a function of temperature for even values of $Q$ between $Q=2$ and $Q=8$ and a $128 \times 128$ lattice. $B = 0$ (purple curves), $B = 1$ (blue curves).}
    \label{thermodynamicswithfield}
\end{figure}

Following with the $128 \times 128$ lattice, its internal energy as a function of $T$ is shown in Fig.~\ref{thermodynamicswithfield}(a), where we can observe that lower values are obtained when $B=1$ (blue curves) than when $B=0$ (purple curves). This is due to the Zeeman term present in the Hamiltonian given by Eq.~(\ref{Clock}), which decreases the ground state energy compared to the case in the absence of a magnetic field. Figure~\ref{thermodynamicswithfield}(b) shows the magnetization $M$ of the system for $Q=2,4, 6,$ and $8$ for $B = 1$ (blue curves) and $B = 0$ (purple curves). First, we notice that all systems start saturated (in a ferromagnetic state), and then $M$ decreases as $T$ increases. The cases with $B=1$ show higher magnetization than those with $B=0$ as $T$ increases for all $Q$ displayed. It is observed that the change in magnetization as a function of temperature happens much faster for larger $Q$ values. This is a direct consequence of the degrees of freedom of the system. For example, we can think of $Q=2$, where two possible orientations for the spin are up and down. In terms of energy, generating a flip of such a configuration is much more difficult. Therefore, the magnetization will be higher for low $Q$ than for higher values of $Q$ at the same temperature. 

Next, we analyze the entropy in Fig.~\ref{thermodynamicswithfield}(c). We observe that for a given $Q$, the entropy is lower when $B = 1$ (blue curves) than for the case of $B = 0$ (purple curves). This is consistent with the interpretation that the magnetic field tends to order the spin system. Also, it is noticeable that for more spin degrees of freedom (as $Q$ increases), the entropy is higher as a function of temperature, which is consistent with the more significant number of accessible states the system acquires as $Q$ increases. These curves show that the entropy fulfills $\lim_{T\rightarrow 0} S(T, B\neq 0) = 0$ because, with even a very low field value, the system at low temperature tends to have only one preferred state due to the presence of the external field applied to the system. It is important to remember that in the case of a null field, the entropy at low temperature has the value of $\lim_{T \rightarrow 0} S(T,0)=\ln Q$. It should be noted that the entropy in Fig.~\ref{thermodynamicswithfield} is normalized to the total number of spins in the lattice, i.e., $S(T,B)/N$. This implies that the quantity $S_{0}/N$ is very small when the field tends to zero (and strictly zero with $B\neq 0$). In the case of lattice size $128 \times 128$ displayed in Fig.~\ref{thermodynamicswithfield}, this number would be given by $\ln(Q)/16384$, in which the most significant value would be provided for $Q=8$, offering a correction of the order of $10^{-4}$.

\section{Magnetocaloric Phenomena} \label{Caloric Phenomena}
\label{Caloric}

To analyze the magnetocaloric effect, we start by treating the total entropy of a material as the sum of three entropies: electronic $S_{e}$, magnetic $S_{m}$, and lattice $S_{l}$.  

\begin{equation}
\label{totalentropyI}
    S(T,B) = S_{e}(T,B) + S_{m}(T,B) + S_{l}(T,B).
\end{equation}

It is essential to mention that the latter equation presumes the separation of the orbital, magnetic, and lattice degrees of freedom. This assumption does not always hold, as evidenced by systems exhibiting Jahn-Teller interactions or magnetoelastic coupling, where these degrees of freedom are intertwined \cite{Englman1970, Kaplan1963, Butterfield2007, Travnikov2021}.


Magnetic entropy $(S_{m})$ strongly depends on the magnetic field. Conversely, it is typically observed that many materials' electronic and lattice entropies are mainly independent of the magnetic field. However, this is not universally applicable; notably, in low-temperature regimes (approximately below 10 K), some studies suggest that the electronic entropy may exhibit nonlinear dependence on the magnetic field \cite{pecharsky2001thermodynamics}. 

In this study, we consider a more conventional system where the magnetic field does not influence the lattice and electronic entropies, allowing us to reformulate Eq.~(\ref{totalentropyI}) as

\begin{equation}
\label{totalentropyII}
    S(T,B) \simeq S_{e}(T) + S_{m}(T,B) + S_{l}(T).
\end{equation}

The quantification of the caloric phenomenon is associated with a thermodynamic refrigerator cycle, where we can take two paths to quantify the effect: i) an adiabatic trajectory and ii) an isothermal trajectory. The temperature variation that suffers the systems along the isentropic process is $\Delta T_{ad}$, and is given by

\begin{equation}
\label{olddeltat}
\Delta T_{ad} = -\int_{B_{i}}^{B_{f}}  \frac{T}{C_{B}} \left(\frac{\partial S}{\partial B}\right)_{T} dB,
\end{equation}
where we use that $C_{B} =\left(\frac{\partial U}{\partial T}\right)_{B} = T \left(\frac{\partial S}{\partial T}\right)_{B}$ corresponds to specific heat at constant $B$. 

In the case of quantifying the effect employing an isothermal trajectory, we obtain an entropy variation at constant temperature, $\Delta S$, given by 

\begin{equation}
\label{deltasisoter}
    \Delta S  = \int_{B_{1}}^{B_{2}} \left(\frac{\partial S}{\partial B}\right)_{T} dB.
\end{equation}

The quantity $\left(\frac{\partial S}{\partial B}\right)_{T} $ that appears in the last two equations can be rewritten in terms of the magnetization of the system via Maxwell's relationship $\left(\frac{\partial S}{\partial B}\right)_{T}=\left(\frac{\partial M}{\partial T}\right)_{B}$.

If we look at Eq.~(\ref{olddeltat}) and Eq.~(\ref{deltasisoter}),  we can find a relationship between these quantities. It is found that $- \Delta S \propto \Delta T_{ad} $. Consequently, it is essential to note that when one has a case in which $-\Delta S > 0$, let's call this kind of response of direct type. The system will heat up, while when the response is $-\Delta S < 0$ type, we call this inverse type response, and consequently, the system will cool down. Therefore, we expect in a direct response a $\Delta T_{ad} > 0$, and for the case of an inverse response, we expect a $\Delta T_{ad} < 0$ for the final result of the caloric phenomena.

For the case of the entropy variation at constant temperature $T$, the magnetocaloric expression can be given by the difference of the entropy at the initial and final point of the process as

\begin{equation}
\label{variationSalpha}
   -\Delta  S \simeq - \Delta S_{m} \simeq S_{m}(T,B_{0})-S_{m}(T,B), 
\end{equation}
where the contributions of $S_{l}$ and $S_{e}$ seen in Eq.\ \ref{totalentropyII} vanish due to their only temperature dependence, and consequently, in an isothermal process (as the one we analyze here), they do not have associated variations as $B$ changes. Equation \ref{variationSalpha} generates a graph of $-\Delta S$ as a function of $T$ for a given $B_{0}$ and final $B$ that quantifies the magnetocaloric phenomena.

For temperature and magnetic field ranges we use in our calculations, we fix the \textit{zero} value of temperature at $T=0.01$, while the smallest field we work is $B=0$. On the other hand, the maximum temperature corresponds to $T=4.02$, and magnetic field of $B=1$, all in $\mathcal{J}$ units.

\section{Results}\label{Resultssec}

\subsection{Direct or indirect caloric response?}

\begin{figure}
	\subfigure[]{
		\includegraphics[width=.50\textwidth]{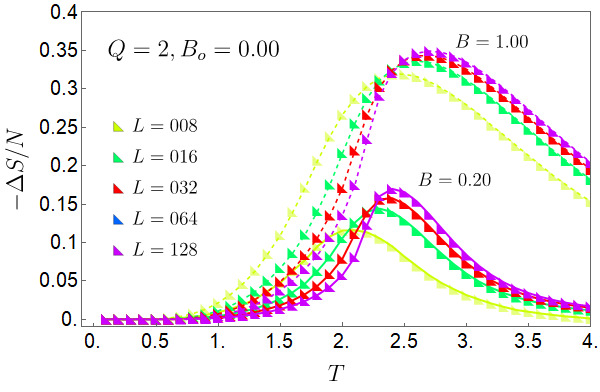}
	}
	\hspace{5mm}
	\subfigure[]{
		\includegraphics[width=.50\textwidth]{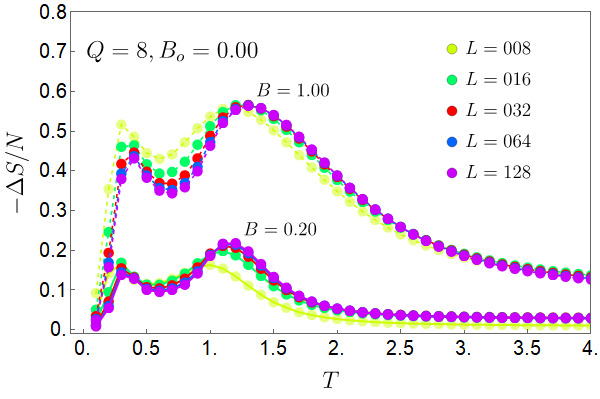}
	}
    \caption{Normalized entropy difference $-\Delta S = S(T, B_{0}=0 ) - S(T, B)$ as a function of temperature for lattices sizes from $8\times 8$ to $128\times 128$ sites, and $B = 0.2$ (lower curves) and $B = 1.0$ (upper curves). (a) $Q=2$, (b) $Q=8$.}
      \label{Response1}
\end{figure}


To understand the caloric response of the system, let us start the analysis with a simple example for $Q=2$, top left panel in Fig.\ \ref{modelfigure}(b), and lattices increasing in size from $8\times 8$ to $128 \times 128$.  The entropy difference that quantifies the MCE in Eq.\ \ref{variationSalpha} is given by $-\Delta S= S(T, 0)-S(T, B)$, where we have selected $B_0=0$. This entropy difference is plotted for different final magnetic fields of $B=0.2$ and $B=1.0$ as a function of the temperature $T$ the process is taken, as shown in Fig.~\ref{Response1}(a). We note that the response of the effect is positive for both presented cases of different $B$, which indicates a direct type behavior for the MCE. It is also observed that for a lattice of $128 \times 128$, and below $T\sim 1.3$, the thermal response of the effect for $B = 0.2$ and $B = 1.0$ is the same. This latter also occurs for smaller lattices but for lower temperatures than $1.3$.

As temperature increases after $T\sim 1.3$, it is observed that the case for $B=1$ is notably different from the one with $B=0.2$, as the peaks for $B=1$ are larger and shifted to higher $T$ for all lattice sizes. The explanation of this phenomenon is relatively simple. The system prefers to be in a ferromagnetic configuration (i.e., in a single configuration state) in an external magnetic field up to reach a maximum in  $-\Delta S$. To remove the spin ordering from this state requires a considerable increase in temperature in terms of energy. This is reflected in the entropy for $Q=2$ and $128\times 128$ lattice for $B = 0$ and $B = 1$, where both entropies are equal up to $T \sim 1.3$, and then they separate as $T$ increases, see two lower curves in Fig.~\ref{thermodynamicswithfield}(c).

Using the same criteria previously discussed, we will obtain similar behavior associated with the sign of $-\Delta S$ for different $Q$ values. That is to say, we will obtain a response of the direct type $-\Delta S > 0$ for $Q=8$ even independent of the lattice size, as can be appreciated in Fig.~\ref{Response1}(b). This is because the entropy without an external magnetic field is always greater than the entropy for any non-zero magnetic field value for any value of $Q$, as we can see for the lattice of $128\times 128$ sites in Fig.~\ref{thermodynamicswithfield}(c). 

In addition, we observe by comparing both panels (a) and (b) of Fig.~\ref{Response1}, that the maxima for $-\Delta S/N$ with $32 \times 32$ and $128 \times 128$ lattice sites do not present significant variations in magnitude neither the $T$ value where they are located in the horizontal axis. This statement is even more noticeable in the case of $Q=8$ in panel (b), indicating that as the lattice size and $Q$ increases, the entropy changes curves tend to be similar for the $T$ range and $B$ values we are considering in these calculations. It is also important to emphasize that an inverse effect $-\Delta S < 0$ can be obtained in an isothermal process, starting from a higher magnetic field and going to a lower magnetic field. Still, it is much more natural to have a system without a magnetic field and activate an external field on the spin system afterward.

For the larger $Q$ value we are considering $Q=8$ in Fig.~\ref{Response1} panel (b), it is observed that the MCE is more significant for the second transition in all presented lattice sizes, as seen by the larger maxima in the temperature range from $T=1$ to $T=1.5$. This is true for large lattices constructed with Monte Carlo simulations and mean-field theory. However, using an exact formulation performed on the canonical ensemble, the opposite is true for small lattices (e.g., a $3\times 3$ lattice, see Fig.\ \ref{deltascompameanfieldexact}(b) for $Q=8$).
Since we are focusing on the thermodynamic limit, we will not delve further into the effect for small lattices in the main text of our work. However, we have added a full Appendix~\ref{MeanField} on exact and mean field calculations for small systems to complement this study.

\subsection{What $Q$ value has the better magnetocaloric response?}

In order to answer which value of $Q$ maximizes the magnetocaloric effect, we have to consider two temperature regimes. Since in our model, temperature is in units of $\mathcal{J}$, it is convenient to think of two temperature regimes, the first one for $0<T<1$, and the second one for $T>1$.

\begin{figure}
\includegraphics[width=.50\textwidth]{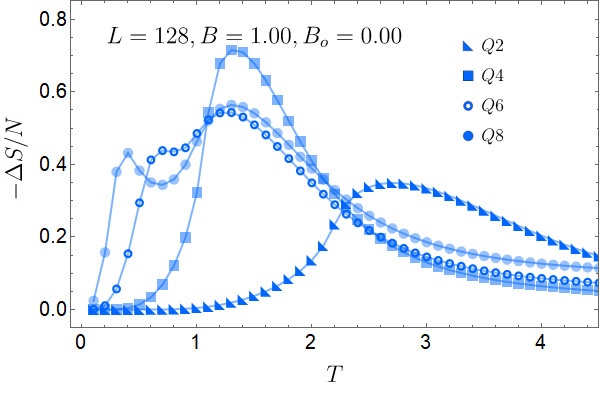}
 \caption{Normalized magnetocaloric effect for a $128 \times 128$ lattice using Monte Carlo simulations for the case of $Q=2$ up to $Q=8$ and external $B=1$.}
\label{qmax}
\end{figure}

From Fig.~\ref{qmax}, we observe the magnetocaloric response for the $128 \times 128$ lattice with even $Q$ values. Due to the double-phase transition in the model for $Q > 5$, each $Q=6$ and $Q=8$ (circle symbols) display two distinct maxima in the caloric effect as a function of $T$. The first maximum (from left to right in $T$) is associated with a FP-BKT phase transition. The second maxima with a BKT-PP transition. For both $Q=6$ and $Q=8$, at temperatures $T < 1$, the magnitudes of $-\Delta S$ are nearly the same, but the first maximum occurs for $Q=8$, reaching its peak at a lower temperature than $Q=6$. This trend aligns with the expected behavior of our model, where increasing the spin degrees of freedom $Q$ allows for similar caloric responses at progressively lower temperatures. As $Q$ increases, the temperature at which the FP-BKT transition occurs shifts leftward, approaching to zero in the theoretical limit as $Q$ tends to infinity. In the higher temperature region $T > 1$, we observe that the caloric effect for $Q=8$ slightly surpasses that of $Q=6$, particularly near the BKT-PP transition. This suggests that, for $Q > 5$, the BKT-PP transition becomes more pronounced with increasing spin degrees of freedom $Q$, amplifying the caloric response as $Q$ grows.

If we only focus on the maximum magnitude of the MCE and not its location in temperature, this is given for $Q=4$ for any lattice size. This is due to the sharp peak observed in the specific heat for $Q=4$ without an external magnetic field, as seen in Fig.~\ref{specificheatImage}(a). If we compare this peak with the one obtained when $B=1$ (Fig.~\ref{specificheatImage}(b)), the magnitude peak decreases almost by half, the most remarkable decrease of all the peaks observed for all the $Q$ studied. This will generate significant entropic differences and a higher associated caloric effect, as shown in Fig.~\ref{qmax}. Opposite is the case for $Q=2$, which has the lowest caloric effect, with its maximum at a higher temperature, thus being the least advantageous for the caloric phenomenology.

\section{Conclusions}\label{Conclusec}


This study investigated the magnetocaloric effect (MCE) in a $Q$-state clock model on a square lattice using Monte Carlo simulations, focusing on spin systems with varying $Q$ orientations. By examining $Q$ values from 2 to 8 across different lattice sizes under free boundary conditions, we quantified the MCE through isothermal changes in the external magnetic field, using the system entropy as a key metric.

Our results demonstrate that the magnetocaloric effect can be significantly enhanced through precise control of the spin degrees of freedom. Specifically, for $ Q \geq 5 $, the system undergoes a double-phase transition characterized by two distinct peaks in specific heat, which correlate with an enhanced caloric response. At lower temperatures ($T < 1 $ ), systems with $ Q = 6 $  and $  Q = 8 $  exhibit a notable caloric effect driven by the ferromagnetic (FP) to Berezinskii–Kosterlitz–Thouless (BKT) transition. Although this transition occurs at different temperatures for each $Q$ value, the resulting caloric response remains nearly identical, indicating that increasing the spin orientations shifts the transition temperature while preserving the effect’s magnitude.

The behavior observed for $ Q = 4 $ is particularly noteworthy, as it presents a single, pronounced peak in the specific heat, maximizing the MCE at a single critical temperature point. Unlike higher $Q$ systems, which involve more complex double-phase transitions, the $  Q = 4 $  case offers an efficient entropy change with simplified thermal control, suggesting its potential utility in applications where streamlined caloric responses are advantageous.

In summary, this study provides a deeper understanding of how spin orientations within the $Q$-state clock model influence the magnetocaloric effect (MCE), offering critical insights for optimizing caloric responses in magnetic refrigeration and other thermodynamic systems. These findings not only advance the theoretical comprehension of phase transitions in multi-state systems but also reveal practical strategies for harnessing these phenomena across diverse domains of  statistical physics such as neural networks, biological systems, and percolation theory. The framework established here opens new pathways for future research into optimizing thermodynamic efficiency in spin-based materials, with promising implications for a range of applied technologies.

\section{Acknowledgments}

  M. A., F. J. P., E. E. V., and P. V.  acknowledge the financial support of ANID Fondecyt (Chile) under contracts 1230055 and 1240582. 
 S. P. A. acknowledges the financial support of UTFSM DPP. F.J.P. acknowledges financial support from “Millennium Nucleus in NanoBioPhysics” project NNBP NCN2021 \textunderscore 021 and UTFSM DGIIE. N. C. acknowledges financial support from ANID Iniciación en Investigación Fondecyt Grant No. 11221088 and DGII-UTA.

\appendix

\section[\appendixname~\thesection]{Exact and Mean Field Approximation }\label{MeanField}

 In this section, we analyze two cases where the MCE effect calculation comes from an entropy that can be derived directly from a canonical partition function $(Z)$. The first corresponds to an exact calculation of all the accessible microstates of the system for a $3 \times 3$ lattice for $Q = 2, 4, 6$ and $8$. In this framework, it is then possible to obtain $Z$ analytically, and hence, all thermodynamic quantities can be derived from it.
 
 The second corresponds to the mean-field theory that assumes that the fluctuations around the average value of the order parameter, in this case, the magnetization per site $\vec{m}=\vec{M}/N$, are so small that they can be neglected. The first term of the Hamiltonian of Eq.~(\ref{Clock}) that corresponds to the interaction term between the spin of the lattice in different sites is modified by performing the following approximations.

We can write the spin term as follows:

\begin{equation}
\label{spinj}
    \vec{S}_{j} = \vec{m} + \delta \vec{S}_{j},
\end{equation}
where $\vec{m}$ is the average thermodynamic spin, the same for all sites in the lattice. Therefore, we have

\begin{equation}
\label{deltaspinj}
   \delta \vec{S}_{j} = \vec{S}_{j} -  \vec{m}_{j}.
\end{equation}

Thus, the spin-spin interaction term can be written as 

\begin{equation}
    \vec{S}_{i} \cdot \vec{S}_{j} = -m^{2} + \vec{m} \cdot \left(\vec{S}_{i} + \vec{S}_{j}\right),
\end{equation}

where we have neglected the square terms of the fluctuation ($O(\delta \vec{S})^{2}$). Therefore, the interaction term of the Hamiltonian of Eq.~(\ref{Clock}) (that we call $\mathcal{H}_{J}$) can take the form

\begin{eqnarray}
\label{interactionaprox}
\nonumber
 \mathcal{H}_{\mathcal{J}} &=& -\mathcal{J}\sum_{\langle i,j \rangle} \left(-m^{2} + \vec{m} \cdot \left(\vec{S}_{i} + \vec{S}_{j}\right)\right) \\ &=& \sum_{i} (2\mathcal{J}m^{2} - \mathcal{J} z \vec{m} \cdot \vec{S}_{i}),
\end{eqnarray}
where now the sum runs for each site in the lattice, and $z$ are the effective nearest neighbors of the model. Consequently, we can define a Hamiltonian per site given by the structure:

\begin{equation}
\label{hamiltonianpersite}
    h_{i} = \frac{z}{2}\mathcal{J}m^{2} - \mathcal{J} z \vec{m}_{i} \cdot \vec{S}_{i} - \vec{B}\cdot \vec{S}_{i}.
\end{equation}

Now, we can calculate the partition function per site, which will depend on $Q$, $B$, $m$ and $T$ given by

\begin{equation}
\label{partitionfunctionfinal}
    Z(q,B,m,T) = \sum_{q} e^{-\frac{\xi(q,B,m)}{T}},
\end{equation}
where $\xi(q,B,m)$ is the energy per site coming from the Hamiltonian of Eq.~(\ref{hamiltonianpersite}).

 To extrapolate the results of mean field approximation of the thermodynamic quantities in smaller lattices, in the Ref.~ \cite{aguilera2022otto}, we discuss the idea of finding an effective nearest neighbor ($z_{eff}$) through a simple optimization protocol. This protocol matches the internal energies obtained for a lattice of  $L\times L$ with $L=3$ employing exact and mean-field calculations. In this way, by releasing the number of neighbors, $z_{eff}\in\,\mathbb{R}^{+}$, and minimizing the internal energy difference (between the exact and approximate case via mean-field), it was found that for $L=3$, the optimal number of neighbors was $z_{eff}=2.67$ (for all values of $Q$).

Amplifying the above qualitatively, we propose an expression for the number of near neighbors effective that adjusts according to the weighting of the effect of non-interacting edges in the system when the lattice has a generic resolution $L\times L$. The effective neighbors for a central spin for the mean field are determined by the following expression (independent of $Q$) for the square lattice with up, down, left, and right near neighbors \cite{aguilera2022otto}

\begin{equation}
\label{zeff}
z_{eff}=\frac{ 4 \times (L-2)^2 + 3 \times 4\times(L-2) + 2 \times 4 }{L^2}.
\end{equation}

Once the partition function of the system has been obtained by either of the two approaches discussed above, it is possible to compute all the thermodynamical observables in a general way through the expressions ( with $k_{B} = 1$ ): 

\begin{equation}
\label{helmholtz}
F=-T \ln Z,
\end{equation}

\begin{equation}
\label{internalenergy}
U=T^{2}\left(\frac{\partial \ln Z}{\partial T}\right)_{B},
\end{equation}
and 
\begin{equation}
\label{specificheat}
C=\left(\frac{\partial U}{\partial T}\right)_{B},
\end{equation}

where $F$ is the Helmholtz free energy, $U$ is the internal energy, and $C$ is the specific heat at the constant magnetic field. In addition, with the differential expression of Helmholtz free energy given by $dF = -S dT- M dB$, we can obtain the entropy and the magnetization of the system given by 

\begin{equation}
\label{entropythermo}
S=-\left(\frac{\partial F}{\partial T}\right)_{B}; \quad M=-\left(\frac{\partial F}{\partial B}\right)_{T}.
 \end{equation}

\begin{figure}
	\subfigure[]{
		\includegraphics[width=.50\textwidth]{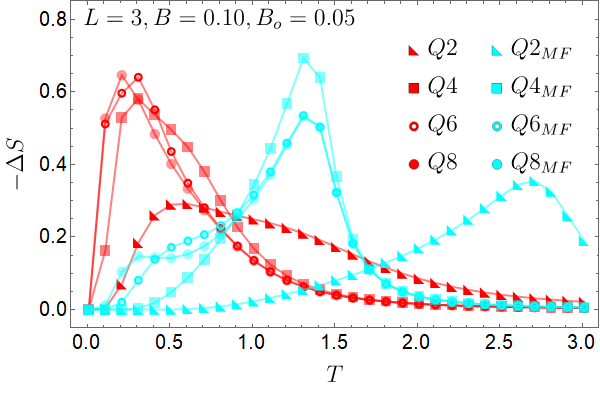}
	}
	\hspace{5mm}
	\subfigure[]{
		\includegraphics[width=.50\textwidth]{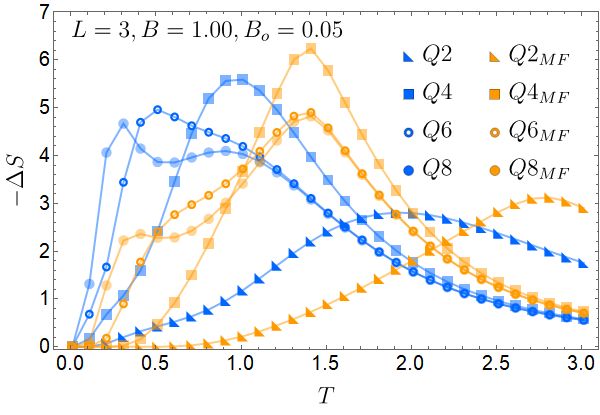}
	}
    \caption{Caloric response for a $3 \times 3$ lattice with an initial field of B=0.05 and a final field of (a) $B=0.1$ and (b) $B=1$ for the exact and mean field cases for $Q=2,4,6$ and $8$.}
    \label{deltascompameanfieldexact}
\end{figure}

\subsection{Small Lattices}

\begin{figure}
\includegraphics[width=0.5\textwidth]{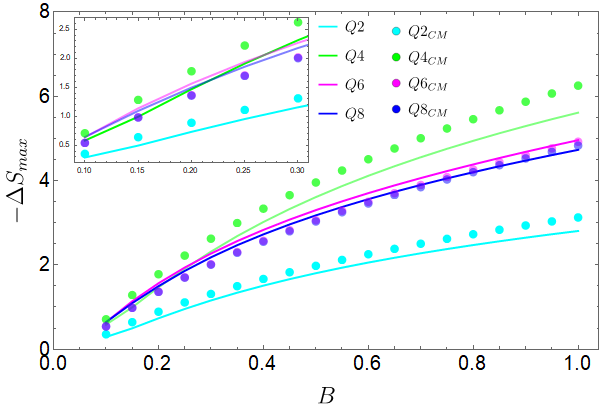}
\caption{Maximum caloric response ($- \Delta S_{max}$) as a function of the applied external field between $0.1$ and $1.0$ for a $3\times 3$ lattice with an initial magnetic field $B = 0.05$ for $Q=2,4,6$ and $8$ employing exact calculations (solid lines) and mean field calculations (dashed lines).}
\label{responsedef}
\end{figure}

We will now analyze the MCE for different values of $Q$ of the $3 \times 3$ lattice, both in the exact case and using the mean field. This is presented in Fig.~\ref{deltascompameanfieldexact}, where the cases for a final field of $B=0.1$ (panel (a)) and the case of $B=1$ (panel (b)) with an initial field of $B=0.05$ are shown. We first notice that for the value of $Q=4$, what was observed for $Q=2$, associated with the location of the maximum temperature effect for both methods, is also true. As the magnetic field increases, the maximum temperature locations become closer. However, a considerable error is obtained when extrapolating mean-field solutions for small lattices, and it has to do with the location of the maximum (or maxima) at $-\Delta S$ when we have the cases $Q=6$ and $Q=8$ as we can appreciate in Fig.~\ref{deltascompameanfieldexact}(b). From this last figure, it can be seen that both $Q = 6$ and $Q = 8$ present two maxima in $-\Delta S$, vestiges of the double transition present for those values of $Q$ in the model, both for the mean field and the exact calculations. We note that the exact calculations indicate that the peak at the lowest temperature is where the most remarkable effect occurs, while the mean field suggests the opposite; the second peak indicates the maximum value of the effect.  In addition, from Fig.~\ref{deltascompameanfieldexact}(a), we observe that in the exact case for the same parameter and lattice configuration, the maximum of the effect is slightly higher for $Q=6$ and $8$ than for $Q=4$. Meanwhile, the mean field indicates that the ordering of the effect from highest to lowest is $Q=4$, $Q=6$, $Q=8$, and finally, $Q=2$.
In contrast, from Fig.~\ref{deltascompameanfieldexact}(b), we can see that the most prominent effect is obtained for $Q=4$ independent of the calculation method. This is remarkable since previous studies found that when applying thermodynamic cycles using the $Q$-clock model as the working system, the maximum efficiency and useful work are obtained for the parameter $Q=4$ \cite{aguilera2022otto}.  These two cases, therefore, indicate that there is a field value at which this inversion observed in the effect maxima occurs. For this purpose, we plot in Fig.~\ref{responsedef} the value of the maximum $-\Delta S_{max}$ as a function of the field independent of the location in temperature at which it occurs both for the mean field and for the exact calculation. From this plot, we can notice that in the exact case, $Q=6$ and $Q=8$, responses higher than $Q=4$ and $Q=2$ are observed up to a field of approximately $\sim 0.26$. After that value, the effect is ordered in the same way as the mean-field results, i.e., in the form:

\begin{eqnarray}
    -\Delta S_{max}^{Q=4} >  -\Delta S_{max}^{Q=6} >  -\Delta S_{max}^{Q=8} > - \Delta S_{max}^{Q=2},
\end{eqnarray}
equation valid only for a final $B > 0.26$.

\begin{figure}
	\subfigure[]{
		\includegraphics[width=.5\textwidth]{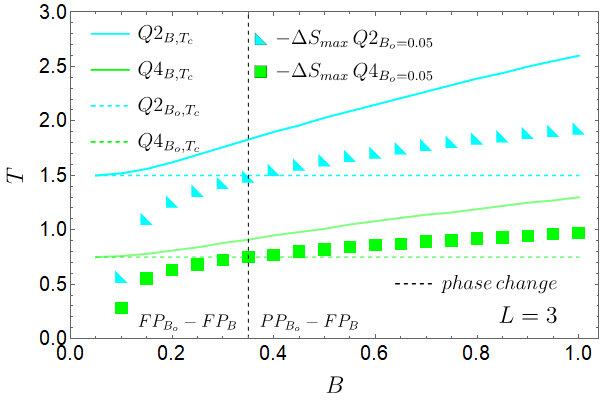}
	}
	\hspace{5mm}
	\subfigure[]{
		\includegraphics[width=.5\textwidth]{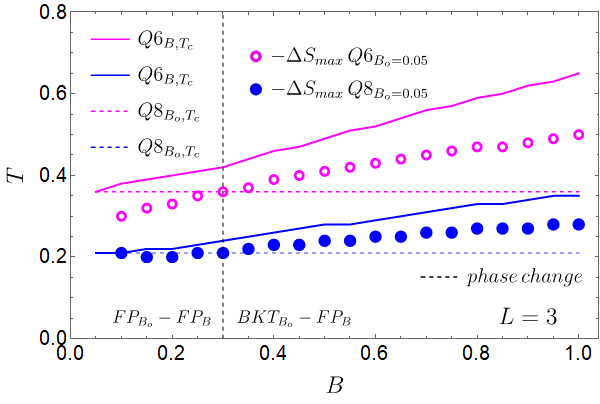}
	}
    \caption{$T$ vs. $B$ phase diagram for (a) $Q = 2$ and $Q = 4$ and for (b) $Q = 6$ and $Q = 8$. The solid lines for (a) represent the critical temperatures of the FP-PP type transition of the final state of the spin system $S(T, B)$, while for (b), the solid lines represent the critical temperatures of the FP-BKT type transition of the final state of the spin system $S(T, B)$. For (a), the light blue triangles indicate the maximum obtained from $-\Delta S= S(T, B_{0}=0.05)- S(T, B)$ with $Q=2$ and the green squares indicate the same but for $Q=4$. The same applies to $Q=6$ represented by magenta circles and $Q=8$ with blue circles. The horizontal dotted lines indicate the critical temperature of the $S(T, B_{0}=0.05)$ state for $Q=2$ (light blue), $Q=4$ (green), $Q=6$ (magenta) and $Q=8$ (blue). The black vertical dotted lines(for panels (a) and (b)) represent the location where the systems maximize $-\Delta S$ passing through an effective phase change.}
    \label{responsedef2}
\end{figure}

From Fig.~\ref{responsedef}, we can also confirm what happens when estimating the maximum effect between the exact and mean-field calculations. Although the mean-field and exact differ quite a bit in the temperature location of the maximum point, if we focus only on the numerical value of effect maximization at $-\Delta S$, the mean-field does not show significant variations compared to the exact calculation even for a small lattice, as shown in this study. The dotted curves in Fig.~\ref{responsedef} representing mean-field calculations differ very little from the solid lines of the exact case, with the most considerable difference observed for $Q=4$ and the minor discrepancies for $Q=6$ and $Q=8$. From this tendency, we can conclude that for values larger than $Q=4$, the numerical estimate of $-\Delta S$ will be closer between the two methods presented. We can say that the mean field is helpful in this context, even for small lattices, in the numerical estimation of the maximum effect. However, it fails logically in the location of the temperature where it occurs because the nature of the approximation used is regularly employed for lattices of larger sizes.

An interesting point to discuss is the location in temperature where the maximum value of the effect is given compared to the critical temperature $(T_{c})$ of the different phases present in the model. This is shown in Fig.~\ref{responsedef2}, which represents a plot of $T$ vs. $B$ for (a) $Q=2$ and $Q=4$ and for (b) $Q=6$ and $Q=8$. To understand these plots, we will start with the cases $Q=2$ and $Q=4$ presented in Fig.~\ref{responsedef2}(a). We must think that the variation of the effect is given by $-\Delta S = S(T, B_{0}=0.05)- S(T, B)$ and hence is described by a spin system starting with an entropy of the form $S(T, B_{0}=0.05)$. This quantity has associated a specific heat $C(T, B_{0}={0.05})$ that defines a critical temperature represented by horizontal lines in Fig~\ref{responsedef2}(a). On the other hand, the continuous lines represent the final entropy state $S(T, B)$, which has an associated specific heat $C(T, B)$, which will deliver as the field changes different critical temperature values, indicating the respective phase transition for these cases. The graphs with geometrical figures (triangular for $Q=2$ and square for $Q=4$) are the points where, for a given field, $- \Delta S$ is the maximum. Precisely, the intersection between the curve with geometric figures and the dotted horizontal one generates, for $Q=2$ and $Q=4$, two possible regions where the system maximizes its effect. Under a field value given by the vertical black line, the system maximizes its effect in a region where the system starts in a ferromagnetic phase $(FP_{B_{0}})$ and ends in a ferromagnetic phase ($FP_{B}$). In contrast, past that critical field value found the system maximizes the caloric response starting from a paramagnetic phase ($PP_{B_{0}}$) and ending in a ferromagnetic phase ($FP_{B}$). The same analysis can be made if we recall that for the exact case of $Q=6$ and $Q=8$, the peak of the effect was concentrated in the first peak of Fig.~\ref{deltascompameanfieldexact} and, therefore, more oriented to transitions relating to the BKT and FP phases.  This is shown in Fig.~\ref{responsedef2}(b), where the maxima are concentrated when the system remains in a ferromagnetic phase or involves transitions linking the BKT and FP phases. We note that the maxima of the effect for $Q=8$ are much closer to the $T_{c}$ of the first phase transition related to an FP-BKT type transition of the final state of the spin system than what occurs for the case of $Q=6$.

%



\end{document}